\documentclass[useAMS,usenatbib]{mn2e}
\usepackage{graphicx}
\usepackage{rotating}
\usepackage{longtable}
\usepackage{lscape}
\def\lsim{\mathrel{\rlap{\lower 3pt \hbox{$\sim$}} \raise 2.0pt \hbox{$<$}}}
\def\gsim{\mathrel{\rlap{\lower 3pt \hbox{$\sim$}} \raise 2.0pt \hbox{$>$}}}
\def\degree{\ensuremath{^\circ}}

\title[A study of six low redshift QSO pairs]
{A study of six low redshift QSO pairs}
\author[Farina et al.]{
	E.~P.~Farina$^{1}$\thanks{E-mail: emanuele.farina@uninsubria.it}, 
        R.~Falomo$^{2}$, and
       	A.~Treves$^{1,3}$\\
       	$^{1}$ Universit\`{a} degli Studi dell'Insubria, via Valleggio 11, I-22100 Como, Italy\\
       	$^{2}$ INAF -- Osservatorio Astronomico di Padova, Vicolo dell'Osservatorio 5, I-35122 Padova, Italy\\
       	$^{3}$ Associated to INAF and INFN\\
	}
	
\begin{document}
\date{ }
\maketitle
\label{firstpage}

\begin{abstract}

	The dynamical properties of six SDSS quasar pairs at $z \leq 0.8$ are 
	investigated. The pairs have proper transverse separation $R_{\perp} 
	\leq 500$~kpc, and velocity difference along the line of sight $\Delta 
	V_r \leq 500$~km/s. If they are bound systems their dynamical mass can 
	be evaluated and compared with that of host galaxies. Evidence 
	is found of an excess of the former mass with respect to the latter. 
	This suggests that these quasar pairs are hosted by galaxies with 
	massive dark halos or that they reside in a group/cluster of galaxies.		

\end{abstract}

\begin{keywords}
QSOs: general
\end{keywords}


\section{Introduction}
	
	Quasars (QSO) are rare and short--lived objects 
	\citep[e.g.,][]{Martini2004,Hopkins2005}, nevertheless a number of 
	associations of QSOs have been discovered in the last decades 
	\citep[e.g.,][]{Shaver1984, Djorgovski1991, ZhdanovSurdej2001}. 
	The study of these systems is important in the understanding of the 
	evolutionary history of galaxies with cosmic time and the mechanism 
	of QSO ignition \citep[e.g.,][]{DiMatteo2005, Foreman2009}. Particular 
	interest has been dedicated to binary QSOs, i.e., two QSOs that reside 
	in the same galaxy and that are characterised by the presence of 
	double systems of emission lines \citep[e.g.,][]{BorosonLauer2009, Rosario2011}. 
	These systems are thought to form in the last stages of a major merger 
	event \citep[e.g.,][and references therein]{ColpiDotti2009}.
	
	The search of QSO pairs (QSOP) at scales from tens to hundreds of 
	kiloparsecs in large surveys was mainly focused on the investigation 
	of QSO clustering properties \citep[e.g.,][]{Hennawi2010, Shen2010} and 
	in particular on the excess, with respect to the large scale extrapolation, 
	found at separations of tens of kiloparsecs  \citep[e.g.,][]{Hennawi2006, Myers2007, Myers2008}. 
	The study of the clustering allows us to estimate the bound mass of the 
	structures inhabited by QSOs \citep[e.g.,][]{Croom2005, Shen2010}, but 
	little attention has been given thus far to the study of the dynamical 
	properties of single QSOPs that, if isolated, are dominated by the mass 
	of their host galaxies~\citep[e.g.,][]{Mortlock1999, Brotherton1999}. 
	Although the Cold Dark Matter models of galaxy formation predict that QSOs, 
	and in particular QSOPs, reside preferentially in particularly rich 
	environments \citep[e.g.,][]{EfstathiouRees1988, Hopkins2008}, some 
	observational evidence shows that QSOPs could be isolated systems 
	\citep[e.g.,][]{Fukugita2004, Boris2007}.
	
	In this paper we look for QSOPs in the Sloan Digital Sky Survey 
	\citep[SDSS;][]{York2000}, with the goal of reconstructing the 
	systemic dynamics of the pairs. We found six QSOPs at redshift~$<0.8$, 
	for which the measurement of [OIII] lines allows us to pursue this 
	study. In~\S\ref{sec:sample} we describe our sample. \S\ref{sec:vel} 
	deals with measurements of radial velocity differences. In~\S\ref{sec:mass} 
	we compute virial masses and compare them with those of the host galaxies. 
	We investigate the QSOP environment in~\S\ref{sec:env}. Implications 
	of our results are discussed in~\S\ref{sec:conclusions}.

	Throughout this paper we consider a concordance cosmology with 
	H$_0=70$~km/s/Mpc, $\Omega_m = 0.3$, and $\Omega_\Lambda=0.7$. 	
	

\section{The quasar pair sample} \label{sec:sample}

	We investigate the catalogue of spectroscopically confirmed QSOs constructed 
	by~\citet{Schneider2010} on the basis of the SDSS DR7 \citep{Abazajian2009} which 
	contains $\sim100,000$ objects. We select as pairs two QSOs that have proper
	transverse separation $R_{\perp} \leq 500$~kpc, and radial velocity difference 
	$\Delta V_r \leq 500$~km/s, as based on SDSS redshifts. $14$ pairs that satisfy 
	the above criteria are found in the redshift range $0.5 \lsim z \lsim 3.3$, with 
	luminosities between $M_V \sim -22$ and $M_V \sim -25$.	
		
	Since we are interested in the dynamical properties of these systems we also 
	require that the forbidden [OIII] lines, which are used to measure the systemic
	velocity of the QSOs (see~\S\ref{sec:vel}), are present in the SDSS spectra. 
	This implies that the candidate QSOPs are at redshifts below 0.8. With this 
	additional condition we obtain a list of six pairs of radio quiet QSOs (see 
	Table~\ref{tab:sample6}), five of them considered also by~\citet{Hennawi2006}.

	\begin{table*}
	\caption{
	Properties of selected QSOPs. 
	$z$ is the redshift from \citet{Schneider2010}, 
	$M_V$ is the absolute magnitude in V--band, 
	$\Delta \theta$ and $R_{\perp}$ are the angular and proper transverse separation, and
	$\Delta V_r$ is the radial velocity difference derived from the redshifts given in the catalogue.
	}\label{tab:sample6}
	\centering
	\begin{tabular}{ll*{10}c}
	\hline
	     &                   &\vline&\multicolumn{2}{c|}{QSO A}&\vline&\multicolumn{2}{c}{QSO B}&\vline&		    &       &                     \\
	ID   & name              &\vline& $z$	 & $M_{V}$ 	   &\vline&  $z$   & $M_{V}$        &\vline& $\Delta\theta$ & $R_{\perp}$  & $\Delta V_r$ \\
	     &                   &\vline&	 & [mag]   	   &\vline&	   & [mag]          &\vline& [arcsec]	    & [kpc] & [km/s]              \\        
	\hline
	QP01 & SDSS J0117+0020AB &\vline& 0.6122 & -22.38  	   &\vline& 0.6130 & -24.65         &\vline& 44           & 300 	  & 149 	  \\
        QP02 & SDSS J0747+4318AB &\vline& 0.5010 & -22.76  	   &\vline& 0.5012 & -22.61         &\vline& \phantom{1}9 & \phantom{1}56 & \phantom{1}40 \\
	QP03 & SDSS J0824+2357AB &\vline& 0.5356 & -23.19  	   &\vline& 0.5365 & -23.19         &\vline& 15           & \phantom{1}94 & 176 	  \\
	QP04 & SDSS J0845+0711AB &\vline& 0.5363 & -23.48  	   &\vline& 0.5373 & -23.20         &\vline& 62           & 393 	  & 195 	  \\
	QP05 & SDSS J0856+5111AB &\vline& 0.5425 & -22.81  	   &\vline& 0.5434 & -23.59         &\vline& 22           & 139 	  & 175 	  \\
	QP06 & SDSS J1249+4719AB &\vline& 0.4375 & -23.09  	   &\vline& 0.4382 & -22.63         &\vline& 79           & 446 	  & 146 	  \\
	\hline			 
	\end{tabular}
	\end{table*}
	
	The probability that they are chance superpositions is rather low. In fact, 
	searching for QSOPs in a random sample generated with the redshift 
	permutation method \citep[e.g.,][]{Osmer1981, ZhdanovSurdej2001}, which consists 
	of maintaining the position of the QSOs fixed, but to randomly permute the redshift, 
	we expect to find $\sim 0.4$ such pairs compared to the $6$ observed. Note that in 
	this new sample most of the correlations between objects are destroyed, but the
	angular correlation between QSOs is preserved, so the result can be considered as 
	an upper limit for the number of chance QSOPs. Thus we assume that all these QSOP 
	are physically associated.
	
\begin{figure}
   	\centering
   	\includegraphics[width=1.\columnwidth]{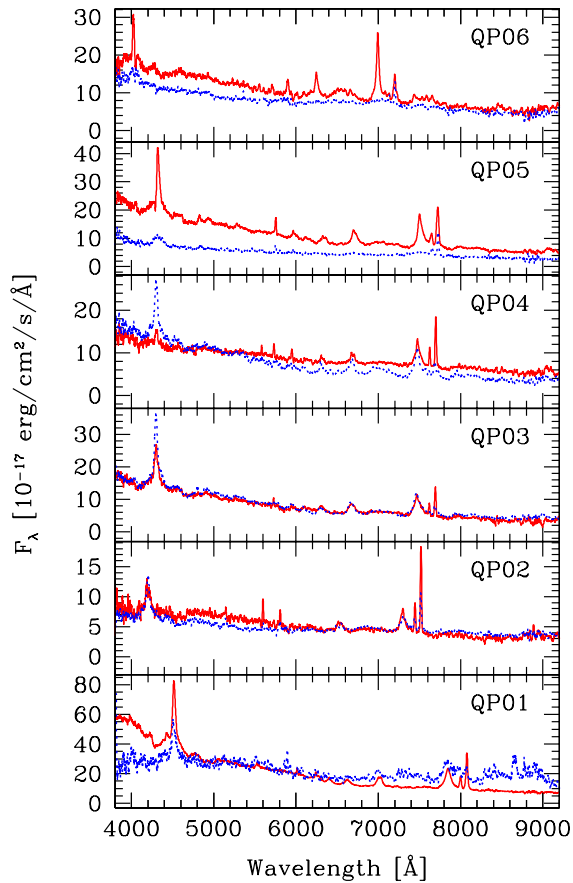}
 	\caption{
	The SDSS spectra of the two QSOs of each pair (background red solid line, foreground blue 
	dotted line). Data are smoothed with a $10$~\AA~boxcar filter. In order to make visible 
	the features of the spectrum of the foreground QSO of QP01, its flux is increased by a factor 
	of 10.
	}\label{fig:2spectra}
    	\end{figure}

	We can exclude that these QSOPs are gravitational lens images because: significant 
	differences in the spectra of the two QSOs are apparent (see Figure~\ref{fig:2spectra}), 
	wide separation ($\Delta \theta>3$~arcsec) lensed QSOs are quite rare \citep{Kochanek1999}, 
	and there is no evidence in SDSS images for luminous galaxies in the foreground of QSOPs 
	that could act as a lens.
	
\section{Velocity differences from [OIII] lines}\label{sec:vel}

	We can also exclude the possibility that the velocity differences can be related to the Hubble 
	Flow and therefore measure the physical distance of the pairs. In fact we verify that, under this 
	hypothesis, in the \citet{Schneider2010} QSO catalogue there are $35$ pairs with 
	$R_{\perp}\leq4$~Mpc and physical radial separation $R_{\parallel}\leq4$~Mpc. Assuming that the 29 
	systems with $R_{\perp}>0.5$~Mpc are homogeneously distributed, we expect $\sim1$ with 
	$R_{\perp}\leq0.5$~kpc, while $6$ are found.
	
	It is well known that the redshifts of QSOs derived from emission lines of various elements can 
	differ by as much as $1,000$~km/s \citep[e.g.,][]{TytlerFan1992, Bonning2007}. Therefore the most 
	reliable estimate of the systemic velocity of the QSOs is obtained from the measurements of narrow 
	forbidden lines, such as [OIII]$_{\lambda 4949}$ and [OIII]$_{\lambda 5007}$ 
	\citep[e.g.,][]{NelsonWhittle1996, Nelson2000, Boroson2005, HewettWild2010}.
	
\begin{figure}
   	\centering
   	\includegraphics[width=1.\columnwidth]{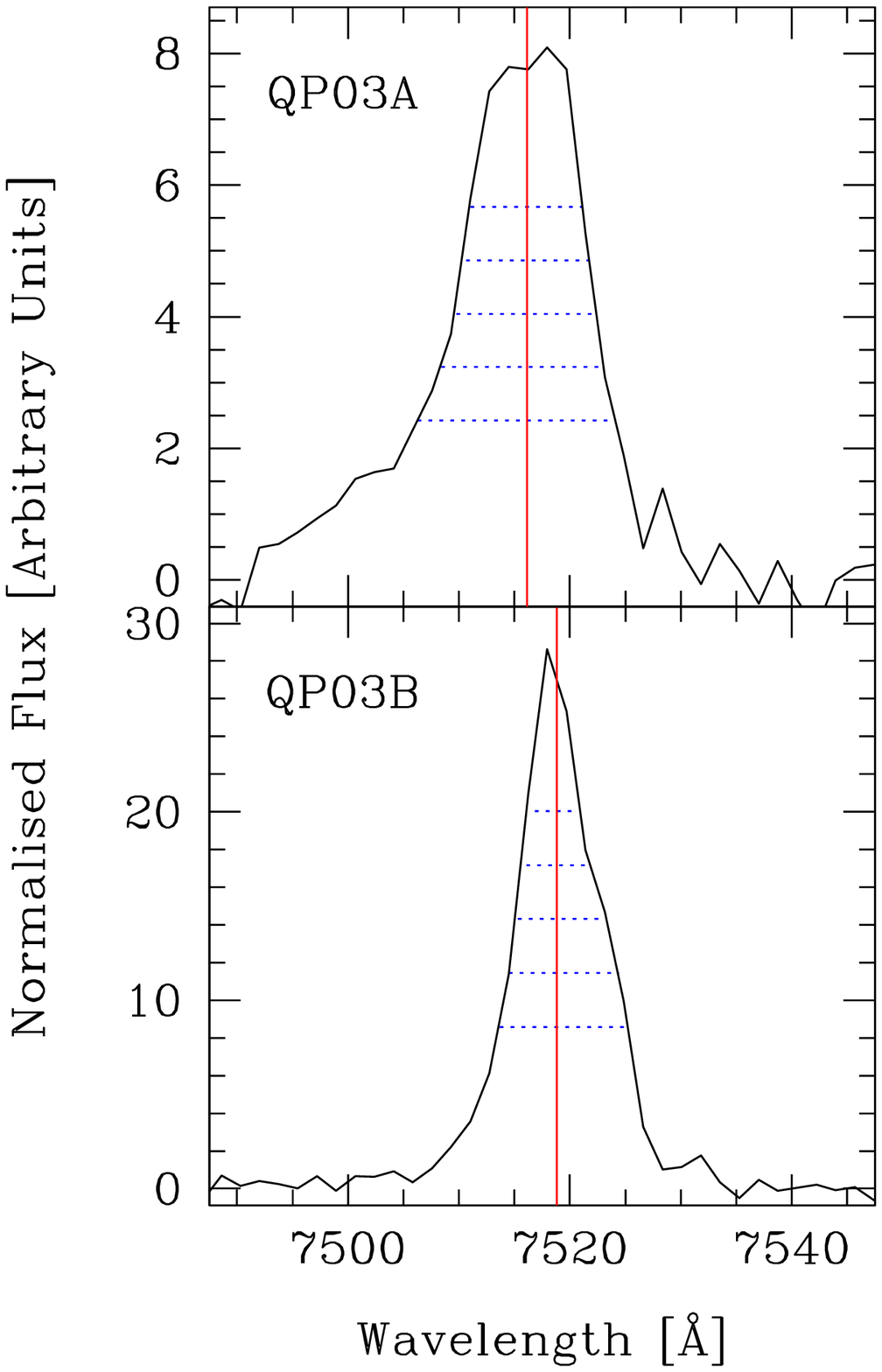}
 	\caption{
	The normalised SDSS spectra of the two QSOs of the pairs QP03 in the region around the emission 
	line [OIII] at \mbox{$\lambda=5007$~\AA}. The central value (red vertical lines) is the median 
	value of the baricentre calculated considering different thresholds of line peak flux (i.e., $30$, 
	$40$, $50$, $60$, and $70$\%, blue dotted horizontal lines).
	}\label{fig:spectra}
    	\end{figure}

	We evaluate the baricentres of the lines considering the flux above various thresholds with respect 
	to the peak flux (see Figure~\ref{fig:spectra}). We take the line position to be the median 
	of the individual measurement of the baricentre, and the corresponding uncertainty 
	is given by their interquartile range. The redshifts and the radial velocity 
	differences that result from these measurements are reported in Table~\ref{tab:red-mass}.
	
	\begin{table*}
	\caption{
	Radial velocity difference and virial mass of the QSOPs.
	$z_n$ is the redshift measured from the [OIII] narrow emission lines;
	$\Delta V_{r}(n)$ is the corresponding radial velocity difference;
	$M_{vir}(min)$ is the minimum virial mass compatible with the uncertainties of 
	the measure of $\Delta V_{r}(n)$.
	}\label{tab:red-mass}
	\centering
	\begin{tabular}{l*{6}c}
	\hline									
	 ID   &\vline& $z_n(A)$ 	   & $z_n(B)$		 &\vline&  $\Delta V_{r}(n)$ & $M_{vir}(min)$		  \\
	      &\vline&  		   &			 &\vline&  [km/s]	   & [$10^{12}\,M_\odot$] \\
	\hline  						
	QP01  &\vline& 0.61142$\pm$0.00078 & 0.61341$\pm$0.00001 &\vline& 370$\pm$171	   	   &  \phantom{1}2.8--20.4	                  \\ 
	QP02  &\vline& 0.50108$\pm$0.00003 & 0.50174$\pm$0.00001 &\vline& 132$\pm$\phantom{10}7	   &  \phantom{1}0.2--\phantom{1}0.3 	                  \\ 
	QP03  &\vline& 0.53527$\pm$0.00009 & 0.53678$\pm$0.00002 &\vline& 295$\pm$\phantom{1}21	   &  \phantom{1}1.6--\phantom{1}2.2 	                  \\ 
	QP04  &\vline& 0.53509$\pm$0.00015 & 0.53754$\pm$0.00002 &\vline& 478$\pm$\phantom{1}35	   &  17.9--24.0	          \\ 
	QP05  &\vline& 0.54322$\pm$0.00003 & 0.54239$\pm$0.00003 &\vline& 161$\pm$\phantom{10}9	   &  \phantom{1}0.7--\phantom{1}0.9 	                  \\ 
	QP06  &\vline& 0.43861$\pm$0.00045 & 0.43859$\pm$0.00001 &\vline& \phantom{1}$4^{+94}_{-4}$&  \phantom{1}\dots  	                  \\ 

	\hline												     	
	\end{tabular}											     					    
	\end{table*}


\section{The mass of QSO pairs}\label{sec:mass}

	Assuming that the QSOPs form bound systems and thus that the velocity difference
	measured is due to the mutual interaction between the two QSOs, we can infer the 
	dynamical mass through the virial theorem:
	\begin{equation}
	M_{vir} = \frac{\Delta V^2 R}{G}
	\end{equation}
	where $\Delta V$ is the relative velocity of the two components, $R$ their separation, 
	and $G$ the gravitational constant. For circular orbits, it is possible to calculate
	the radial component of the relative velocity ($\Delta V_{r}$) from the redshift difference 
	($\Delta z$). One has:
	\begin{equation}
	M_{vir} = C \left(\frac{c\,\Delta z}{1+z}\right)^2\frac{R_{\perp}}{G}
	\label{eq:mpair}
	\end{equation}
	where $c$ is the speed of light, $R_{\perp}$ the proper transverse separation of the pair (see 
	Table~\ref{tab:sample6}), and the factor $C$ depends only on the inclination angle of the orbital 
	plane $\iota$, and on the phase angle $\phi$ and is given by:
	\begin{equation}
	C^{-1} = \left(\sin \phi \sin \iota  \right)^2 \times \sqrt{\sin^2 \phi + \cos^2\phi \cos^2 \iota}
	\end{equation}	
	The average values of $C$ is $\langle C \rangle=3.4$ and the minimum value is $C_{min}=1$.
	
	In Table~\ref{tab:red-mass} we report for each QSOP the minimum virial mass ($M_{vir}(min)$, 
	corresponding to $C=1$), which represents the 
	minimum mass of the system to be bound. In the case 
	of QP06, since there is no significant difference of radial velocity, we cannot estimate its virial 
	mass. In this case we are probably observing the pair orbit nearly face on.  

	It is of interest to compare these $M_{vir}(min)$ with the expected total mass of the pair based 
	on the mass of their host galaxies. According to 
	available measurements of QSO host galaxies 
	\citep[e.g.,][and references therein]{Kotilainen2009} it is found that their mass changes little with 
	redshift. The typical range of host mass, based on the galactic luminosity, for objects at $z<1$ is 
	$\sim 0.3-1.3 \times 10^{12} M_{\odot}$ \citep[][and references therein]{Decarli2010}.
	
	While for three QSO pairs (QP02, QP03, and QP05)  their $M_{vir}(min)$ is consistent with that expected 
	by the typical host galaxy masses, in two cases (QP01 and QP04) the minimum virial mass is substantially 
	larger than that of their host galaxies (see Table~\ref{tab:red-mass}). If one assumes the average value 
	of $C$ ($\langle C \rangle=3.4$) instead of its minimum, then the above cases are further strengthened and 
	also QP03 would exhibit a significant mass excess. For the whole (small) sample the median value for 
	the $M_{vir}$ is $6.5\times 10^{12} M_{\odot}$.

	A possible explanation for this mass excess is that QP01 and QP04 belong to a group or a cluster of 
	galaxies. In this case in fact the measured velocity difference depends on the overall mass distribution. 
	In the next session we investigate this possibility.

	
\section{QSO pairs' environment}\label{sec:env}
	
	We searched the SDSS $i$--band images for an overdensity of galaxies that could justify the mass excess 
	discussed above. The SDSS magnitude limit in this band is $21.3$~mag \citep{York2000}, thus it allows us 
	to reach $\sim ( M^* + 1 )$, where $M^* = -20.5$ at $z = 0.5$ \citep[][]{Wolf2003}, therefore these images 
	permit us to detect only the bright part of the galaxy luminosity function.

	The galaxy search was performed using SExtractor \citep{Bertin1996} on the SDSS images in an area of $4$~Mpc 
	around each pair. The threshold limits for the detections is set at $1.5$ times over the {\it rms} of the 
	background, and we classified as {\it galaxy} all the sources with the STARCLASS parameter lower than $0.2$. 
	The number of galaxies in the fields (see Table~\ref{tab:ML}) is consistent 
	with the expectation from the study performed up to $I=24$~mag by \citet{Postman1998} on a region of 
	$4\degree\times4\degree$, and the number of galactic stars with the prediction
	of the \verb"TRILEGAL" package\footnote{{\rm http://stev.oapd.inaf.it/cgi-bin/trilegal\_1.4}} by 
	\citet{Girardi2005}.
	
	\begin{table}
	\caption{Environment of QSOPs.
	n(bkg) is the density of galaxies in the region between $2$~Mpc and $4$~Mpc, 
	N($<0.5$~Mpc) is the number of galaxies in the inner $500$~kpc, and
	n($<0.5$~Mpc) is the corresponding density.	 
	$M/L$ is the minimum mass--to--light ratio that could have a galaxy cluster detected on SDSS i--band 
	images ($3\sigma$ over the background).	
	The associated uncertainties represent the $1\sigma$ statistical fluctuations.
	}\label{tab:ML}
	\centering
	\begin{tabular}{lcccc}
	\hline									
	 ID   & n(bkg)           & N($<0.5$~Mpc) & n($<0.5$~Mpc)   & $M/L$  \\
	      & [arcmin$^{-2}$]  &		  & [arcmin$^{-2}$]  &	      \\
	\hline   					 
	QP01  & 1.2$\pm$0.2 & \phantom{1}5   	& 1.4$\pm$0.6 & $\phantom{1}\gsim30 $ \\ 
	QP02  & 1.4$\pm$0.1 & \phantom{1}8   	& 1.4$\pm$0.5 & $\phantom{10}\gsim5  $ \\ 
	QP03  & 2.1$\pm$0.1 & 26  		& 4.7$\pm$0.9 & $\phantom{10}\gsim2  $ \\ 
	QP04  & 1.9$\pm$0.1 & 15  		& 2.7$\pm$0.7 & $\gsim100$ \\ 
	QP05  & 1.8$\pm$0.2 & \phantom{1}5   	& 0.9$\pm$0.4 & $\phantom{10\gsim}5  $ \\ 
	QP06  & 1.9$\pm$0.2 & 12  		& 2.3$\pm$0.7 & \dots	     \\ 
	\hline												     	
	\end{tabular}											     					    
	\end{table}
		
	In order to highlight a possible overdensity around the QSOPs, we compute the number of galaxies in annuli 
	of $500$~kpc radius, starting from the centre of each pair. We then compare the galaxy density in the first 
	$500$~kpc with that in the region between $2$~Mpc and $4$~Mpc, assumed as background. These values are 
	reported in Table~\ref{tab:ML}.	 
	Only QP03 shows a significant overdensity of galaxies. In the other cases there is no evidence for a galaxy
	excess above the background by more than $3\sigma$.
	
	We evaluate the expected density of galaxies brighter than the SDSS luminosity limits ($i\sim21.3$~mag) if a 
	cluster of mass $M_{tot}=M_{vir}(min)$ were associated 
	with the QSOPs. We assume that the galaxies of the cluster 
	follow the Schechter luminosity function with parameters given by~\citet{Wolf2003}, and that the galaxies
	are distributed according to a King profile with a virial radius calculated from the virial mass following 
	the relations reported by~\citet{Girardi1998}.	
	
\begin{figure}
   	\centering
   	\includegraphics[width=1.\columnwidth]{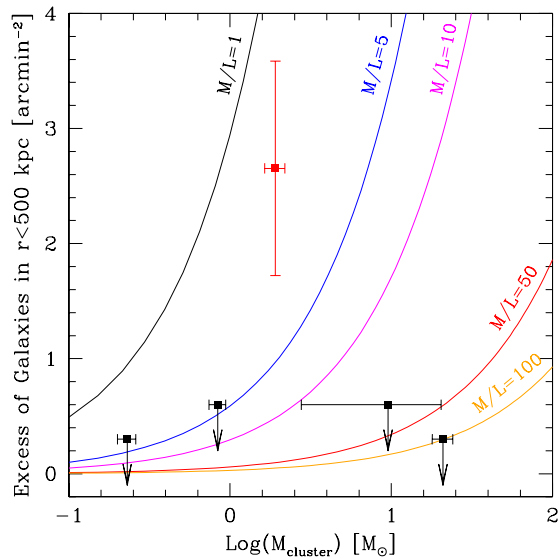}
 	\caption{
	Expected central density of galaxies in SDSS i--band images as a function of the mass of the clusters 
	for various value of the mass--to--light ratio (filled lines). Black square points are the QSOPs for 
	which no significant evidences of an overdensity of galaxies are present. For these systems we assume 
	that, if a cluster is present, it must have a density $<3\sigma$  the variation of the background. 
	The red square point indicates the pair QP03, for which SDSS images highlight a significant overdensity
	of galaxies in the first $500$~kpc from the QSOP.}\label{fig:expclus}
    	\end{figure}

	We compare the expected galaxy density with that observed in SDSS images (see Figure~\ref{fig:expclus}). 
	In all cases but one,we do not find indications for overdensities larger than $3$ times the variation of 
	the background, thus, to explain the minimum virial masses of the pairs, these systems require a 
	mass--to--light ratio $M/L\gsim5$--$100$ (see Table~\ref{tab:ML}). Note that these values are comparable with
	those reported in various studies on dynamical properties of galaxy clusters 
	\citep[e.g.,][and references therein]{Popesso2005}.
	
\section{Conclusions}\label{sec:conclusions}

	The analysis of the properties of 6 low redshift QSO pairs has shown that in at least two cases the 
	dynamical mass of the pair exceeds, by a factor $\gsim 10$, that expected from their host galaxies. 	
	A possible explanation of this excess is that the QSO host galaxies are surrounded by dark matter 
	halos with masses similar to those found in massive ellipticals \citep[e.g.,][]{Napolitano2009}. 
	Alternatively the observed velocity differences could be due to the presence of a cluster or a group of 
	galaxies associated with the QSO pairs. An analysis of SDSS i--band images shows evidence for a significant 
	overdensity of galaxies in only one case. For the other systems a lower limit to the mass--to--light 
	ratio was determined at $M/L\gsim5$--$100$ for galaxy clusters with masses equal to the virial masses 
	of the pairs.
	
	In order to strengthen the evidence of a mass excess, we can consider a larger sample given by
	the lists of already known QSO pairs \citep{Schneider2010, Hennawi2006, Hennawi2010, Myers2008}. Most of 
	these systems are at $z > 0.8$, excellent instrument capabilities are thus required to perform these studies.


\section*{Acknowledgements}
      
	We acknowledge helpful discussions with R.~Decarli, R.~Rampazzo, and M.~Clemens. For this work EPF was
	supported by Societ{\`a} Carlo Gavazzi S.p.A. and by Thales Alenia Space Italia S.p.A.
	
	Data for this work are from the Sloan Digital Sky Survey. The SDSS Web 
	Site is {\texttt http://www.sdss.org/}.


\label{lastpage}

\end{document}